\documentclass[review]{elsarticle}

\usepackage{lineno,hyperref}
\modulolinenumbers[5]

\journal{Journal of \LaTeX\ Templates}

\begin{document}

\begin{frontmatter}

\title{Performance of TPC detector prototype integrated with UV laser tracks for the circular collider}

\author[c,b,a]{Z.Y. Yuan}
\author[a,b,c]{H.R. Qi}
\cortext[mycorrespondingauthor]{Correspondence to: 19B Yuquan Road, Shijingshan District, Beijing 100049, China. Email address: qihr@ihep.ac.cn (H.R. Qi).}
\author[d]{Y. Chang}
\author[e]{L.W. Yu}
\author[f]{Y.M. Cai}
\author[a,b,c]{H.Y. Zhang}
\author[a,b,c]{J. Zhang}
\author[a,b,c]{Q. Ouyang}
\author[f]{Y.L. Li}
\author[f]{Z. Deng}
\author[f]{H. Gong}

\address[a]{Institute of High Energy Physics, Chinese Academy of Sciences, Beijing 100049 China}
\address[b]{State Key Laboratory of Particle Detection and Electronics, Beijing 100049, China}
\address[c]{University of Chinese Academy of Sciences, Beijing 100049, China}
\address[d]{Nankai University, Tianjin 300071, China}
\address[e]{Liaoning University, Liaoning 110036, China}
\address[f]{Department of Engineering Physics, Tsinghua University, Beijing 100084, China}

\begin{abstract}
Several new experimental concepts in high-energy particle physics have been proposed in recent years. The physical goals include precisely measuring the properties of particles such as Higgs, Z and W, and even looking for signs of new physics at future colliders. To meet the evolving requirements for particle track detector, Time Projection Chamber(TPC) detector prototype integrated with a UV laser track system was developed for the main track detector at Circular Electron Positron Collider(CEPC). This prototype consists of 6 horizontal laser tracks around TPC detector chamber, a fast electronics readout of 1280 channels, a GEM detector with $200\times 200\,mm^2$ active area, and the DAQ system. The hit resolution, dE/dx resolution and drift velocity were studied by measuring and analyzing using the TPC prototype and UV laser tracks. The dE/dx resolution of the prototype was measured to be $(8.9\pm0.4)\,\%$. Extrapolating this to CEPC TPC with 220 layers and longer track, the resolution was estimated to be $(3.36\pm0.26)\,\%$. All results indicated that the TPC detector prototype integrated with UV laser tracks can work well.
\end{abstract}


\end{frontmatter}

\linenumbers

\section{Introduction}
With the discovery of the Higgs boson at the Large Hadron Collider(LHC), the feasibility of future colliders such as the International Linear Collider(ILC)\cite{list2014international}, CEPC\cite{cepc2018cepc}, and the Future Circular Collider(FCC-ee)\cite{fcc2019fcc} are investigated with higher performance requirements. CEPC was designed as a high-luminosity and high-center-of-mass energy collider project. It adopts a double-ring structure layout and has two colliding points. The ring circumference is about $100\, km$. CEPC can work as the W and Z factory with a center-of-mass energy of about $160\, GeV$ and $91\, GeV$, for detailed research of related parameters in the standard model. It can also work in a way with higher center-of-mass energy (about $240\, GeV$) to measure the properties of Higgs boson. Some of the research goals of CEPC are similar to ILC.

Three detector concept options were proposed for CEPC detector system. The baseline design concept uses TPC as the primary track detector under the high magnetic field. Other options include an all-silicon detector concept, and an Innovative Detector for Electron-positron Accelerators(IDEA) detector concept using drift chambers inside 2 T magnetic field. For the main track detector TPC of CEPC, the physics requirements would be met to very high momentum and position resolutions. Its track momentum resolution need reach to $\Delta (1/p_{T}) \sim 10^{-4}\, GeV^{-1}$, and the position resolutions of $\sigma _{r\phi}$ and $\sigma _{z}$ reach to $100\,\mu m$ and $500\,\mu m$, respectively. In addition, the dE/dx resolution was required to be better than $5\%$ in the CEPC CDR\cite{cepc2018cepc}.

For traditional TPC, dE/dx resolution performance can reach $6.8\%$ and $5.0\%$ for STAR\cite{xu2010improving} and ALICE TPC\cite{alme2010alice}, respectively. In recent years, with the development of the pixelized readout, a GridPixel TPC\cite{ligtenberg2018performance} based on the Timepix3 chip is developed for future colliders. Its performance is studied in a beam test with $2.5\, GeV$ electrons, and its dE/dx resolution reached $4.1\%$ by the truncated sum. 

For the convenience of TPC performance test, laser beam was used to simulate the particle tracks instead of electron beams with specific energy. It has several advantages, such as minor ionization density fluctuations, easy control and a better spatial resolution in long-track and double-track measurements. In addition, the laser beam does not influence or influenced by the electromagnetic field, so it can be used to study the influence of the electromagnetic field on tracks.

In this paper, some performance and test results of TPC prototype integrated with UV laser tracks were presented including the hit resolution, dE/dx resolution, and drift velocity.

\section{Experiment Setup}
\begin{figure}[htbp]
	\begin{center}
		\includegraphics[width=0.76\linewidth]{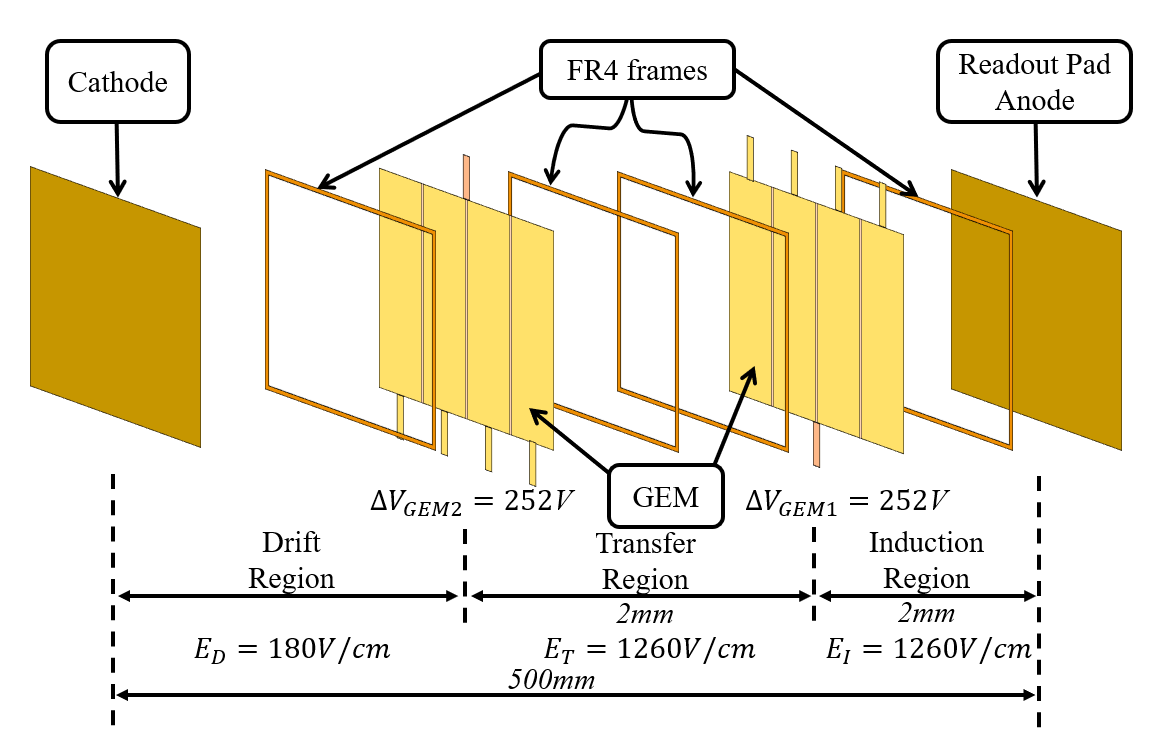}
		\caption{Diagram of the detector readout structure.} 
		\label{Fig1-1}
	\end{center}
\end{figure}

Innovatively, $266\,nm$ ultraviolet laser is used for the specific track reconstruction, performance measurement, and operating condition monitoring. The TPC prototype with 6 horizontal laser tracks is designed and optimized, with a drift length of $500\,mm$, and an effective area of $200\,mm\times 200\,mm$ for double GEM readout structure, as shown in Fig.~\ref{Fig1-1}. The working gas is T2K gas mixture($Ar/CF_{4}/iC_{4}H_{10}=95/3/2$) and the operation gain is about 3400. 

\begin{figure}[htbp]
	\begin{center}
		\includegraphics[width=0.76\linewidth]{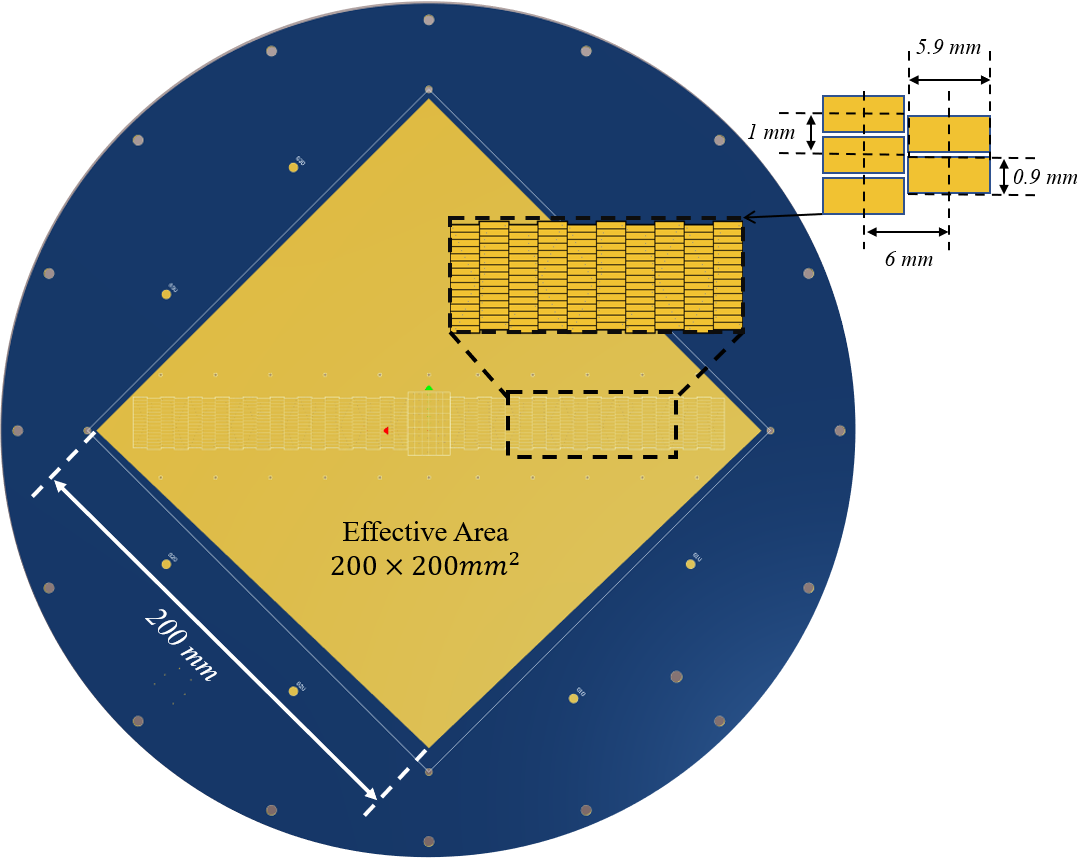}   
		\caption{The layout diagram of readout pads.} 
		\label{Fig1-2}
	\end{center}
\end{figure}

TPC readout pads are arranged along the horizontal laser beam tracks, to record charge on the ground plane. To allow the primary charge to be absorbed by the ground plane as much as possible during the electrons diffusion process, the pads are designed to perpendicular to the laser direction, while in the absence of magnetic field, the maximum drift length of electron diffusion is $500\,mm$. As shown in Fig.~\ref{Fig1-2}, about 700 pads are arranged along the longest diagonal of the square area as readout structures to make full use of the effective detection area. For such a design, one reconstruction track could reach to 38 hit points. Each pad is $1 \times 6\,mm^2$ in area, and 15 pads in each column. Half interleaved columns are set to reduce the error of adjacent sampling.

The Q-smart 100 model Nd-YAG laser of Quantel Company was used in the experiment. After quadrupling the frequency, the wavelength is adjusted to $266\,nm$. The output laser is a $TEM_{00}$ Gaussian beam with a spot diameter of about $4.55\,mm$ and a divergence angle of $0.52\,mrad$. The laser host can adjust the laser energy with a maximum output energy of $20\,mJ$/pulse and a minimum energy of $50\,\mu J$/pulse.

\begin{figure}[htbp]
	\begin{center}
		\includegraphics[width=0.9\linewidth]{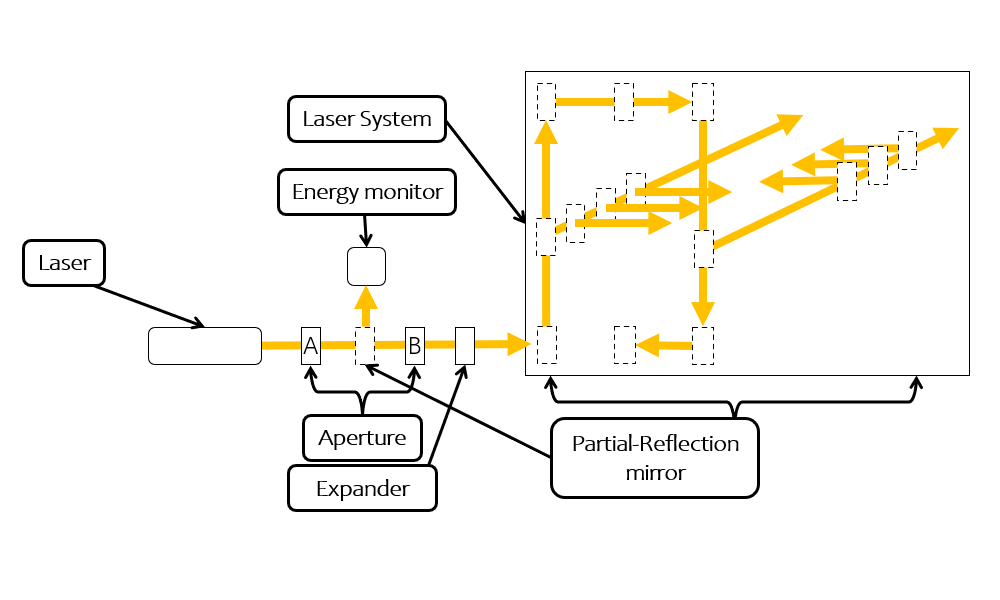}
		\caption{266nm UV laser mimic setup for the UV laser optical transmission(Left) and UV laser mapping system(Right).} 
		\label{Fig3}
	\end{center}
\end{figure}

\begin{figure}[htbp]
	\begin{center}
		\includegraphics[width=0.76\linewidth]{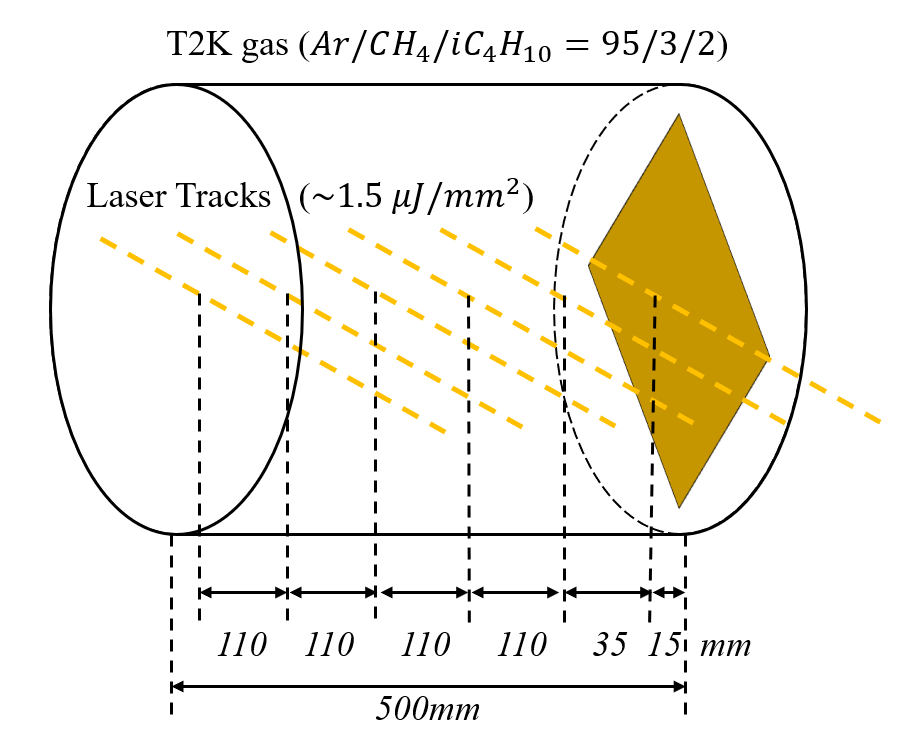}
		\caption{The sketch of the detector with laser tracks.} 
		\label{Fig2}
	\end{center}
\end{figure}

In the optical system shown in Fig.~\ref{Fig3}, the laser first passes through aperture-A to remove the $532\,nm$ UV laser part. Subsequently, it is split by a semi-transmissive mirror($50\,\%$ transmission and $50\,\%$ reflection). One beam is monitored by the StarLite energy meter of Ophir. The primary UV laser beam is transmitted through apertures A and B as collimators, so that the laser beam spot diameter is kept at $0.8\,mm$. Finally, it enters the laser system through a beam expander device to produce the specific splitting UV laser beam tracks\cite{hai2019signal}. The laser energy density of each track is about $1.5\,\mu J/mm^2$ and the relative position is shown in the Fig.~\ref{Fig2}. 

\section{Event selection and laser track reconstruction}
\subsection{Event selection}
The performance of detector is evaluated using filtered events with clean tracks. Amounts of laser track events are collected and optimized, to analyze and determine the final reconstructed tracks.

The energy of laser pulses will fluctuate to a certain extent. The spectrum of laser energy loss due to ionization will be a Landau-like distribution instead of a Gaussian distribution. So the events are chosen with energies within the range $E_{mean} \pm \sigma$. This selection keeps $68.3\%$ of the events and makes the ionization energy loss almost Gaussian distribution. $E_{mean}$ and $\sigma$ are the mean value and standard deviation of the laser energy distribution, respectively. For the TPC detector, four layers of laser track are included in an event, and their ranges of the Time of Arrival(ToA) are shown in Table.~\ref{Tab1}. A hit is accepted with at least two pads in each column triggered. The events are one-to-one with the laser energies. If the energy of the event is within the selected range, the event is defined as a match. The summary of the event selection cuts is given in Table.~\ref{Tab1}.

\begin{table}[htbp]
	\begin{center}
		\caption{Summary of the event selection cuts.}
		\label{Tab1}
		\begin{tabular}{ p{3.5cm} p{3.5cm} p{3.5cm} }
			\hline
			\hline
			Laser energy monitor & Variation range & $E_{mean} \pm \sigma$ \\
			\hline
			TPC detector & Hit ToA & layer$\#1$  $2.6\sim2.9\,\mu s$\\
			&&layer$\#2$ $5.7\sim6.0\,\mu s$\\
			&&layer$\#3$ $8.2\sim8.5\,\mu s$\\
			&&layer$\#4$ $10.5\sim11.0\,\mu s$\\
			& Trigger pads & $\ge 2$ for each column\\
			\hline
			Laser and detector & \multicolumn{2}{l}{The laser control chassis triggers the energy monitor}\\
			& \multicolumn{2}{l}{and DAQ system at the same time. } \\
			\hline
			\hline
		\end{tabular}
	\end{center}
\end{table}

\subsection{Laser track reconstruction}
\begin{figure}[htbp]
	\begin{center}
		\includegraphics[width=0.9\linewidth]{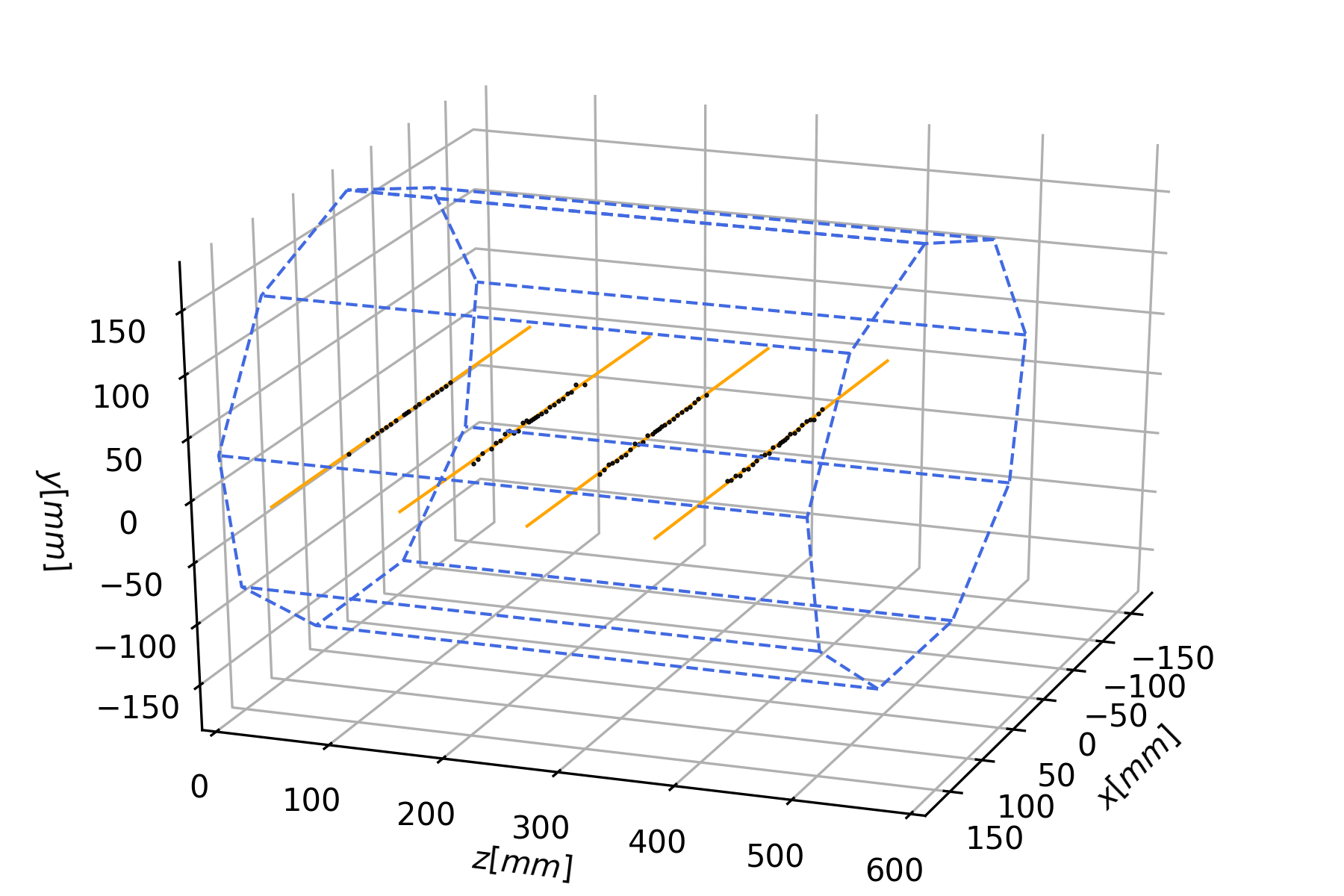}
		\caption{A reconstruction event with 26 hits for each layer and the track fit.} 
		\label{Fig4}
	\end{center}
\end{figure}

Four horizontal laser pulses track along the x-axis direction through the Quartz glass window from the outer TPC chamber to the inner at the specific drift position. The direction of the drift field is the same as the positive z-axis. The readout plane is located on the xy plane with z=0. Four among the 6 laser tracks were chosen to reconstruct the events, avoiding electric field distortion at the edges of the barrel. The four laser tracks are at $z=50$, $160$, $270$, and $380\,mm$. The xy positions of the primary electron ionization is obtained by the charge spread on the pads, and its z position is given by the drift time. Tracks are fitted using singular value decomposition(SVD), a widely used algorithm in machine learning\cite{mandel1982use}. The average position of hits is calculated, and then SVD is performed on the residual centered on it to obtain the direction vector of the track. An example of hits with fitted tracks are shown in Fig.~\ref{Fig4}.

\section{Performance results}
\subsection{Hit resolution}
\begin{figure}[htbp]
	\begin{center}
		\includegraphics[width=0.9\linewidth]{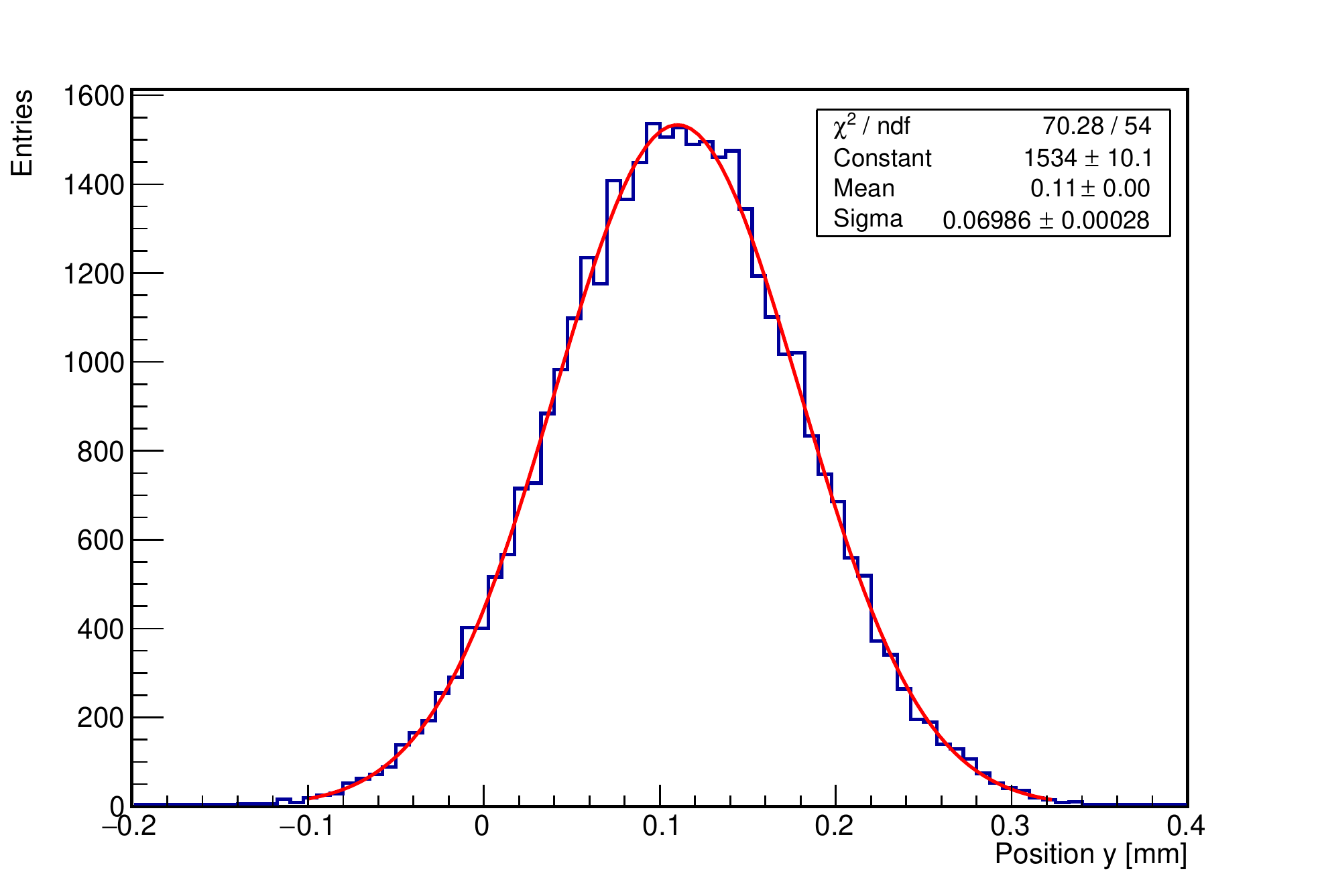}
		\caption{Distribution of the position y at $z = 160\, mm$. A Gaussian fit was applied in a $3\sigma$-range around the mean. The $\chi ^2/ndf = 70.28/54$ of the fit reflects the slight asymmetry visible in the distribution.} 
		\label{Fig5}
	\end{center}
\end{figure}

The TPC hit position resolution measures the accuracy of the track reconstruction on the one hand, and directly affects the momentum resolution on the other hand, so it's also one of the key properties of the TPC. The position resolution of the hit is determined by the transverse diffusion coefficient $D_{T}$ and the drift distance $z-z_{0}$ of the transverse drift part, the pad size $h$ and the laser width $w$, and the effective number of electrons $N_{eff}$ which depends on the energy density of the laser. The position resolution $\sigma_{y}$\cite{ligtenberg2018performance,blum2008particle} is given below:
\begin{equation}
	\sigma_{y}^{2}=\frac{D_{T}^{2}}{N_{eff}}(z-z_{0})+\frac{h^{2}}{12 N_{eff}}+\frac{w^{2}}{12 N_{eff}},
	\label{Eq1}
\end{equation}
where $z_{0}$ is the position of the upper surface of the first GEM layer. There is a power law relation\cite{cai2020investigation} between the laser energy density and its ionization density, given in Eq.~\ref{Eq2}.
\begin{equation}
	I(E)=I_{0} E^{2.27},
	\label{Eq2}
\end{equation}
where $I$ is the laser ionisation density which is the number of primary electrons per centimeter, $I_{0}$ equals to the laser ionisation density of a minimum ionization particle(MIP), and $E$ is the laser energy density. The distribution of the position y with $\sigma_y$ equal to $(69.9\pm 0.3)\, \mu m$ is shown in Figure.~\ref{Fig5}.

\begin{figure}[htbp]
	\begin{center}
		\includegraphics[width=0.9\linewidth]{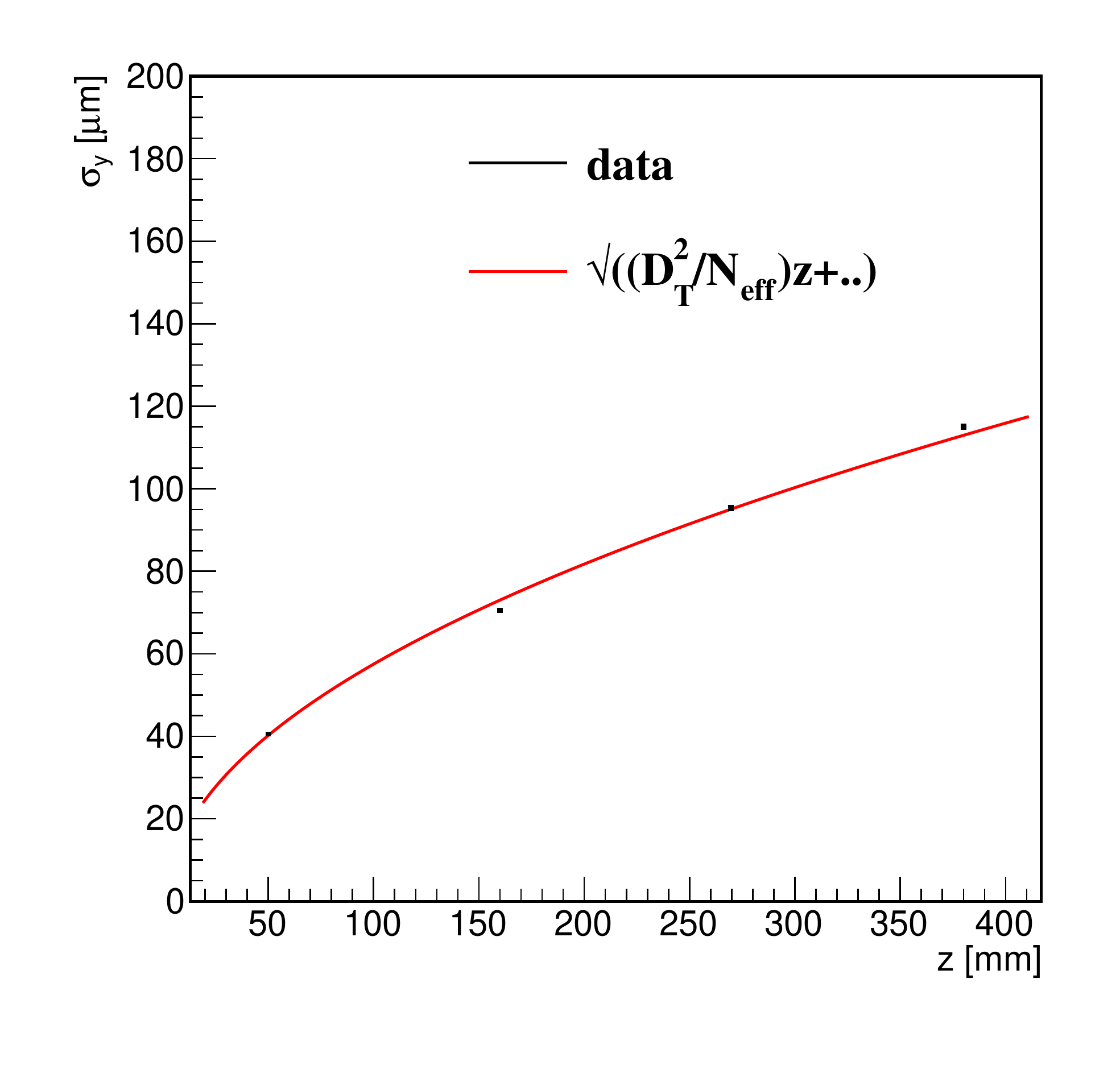}
		\caption{Measured hit resolution $\sigma_{y}$(black points) as a function of drift length $z$ fitted with Eq.~\ref{Eq1} (red line).} 
		\label{Fig6}
	\end{center}
\end{figure}

For different drift distances(layers of the laser track), the measured hit resolution as a function of drift length $z$ is shown in Fig.~\ref{Fig6}. The measured transverse diffusion coefficient $D_{T}$ is $(310.7\pm 7.6)\, \mu m/\sqrt{cm}$. The transverse diffusion coefficient measured by varying the drift distance is similar to the expected value $312.7\, \mu m/\sqrt{cm}$ simulated by Garfield++\cite{veenhof1998garfield}.

\subsection{Energy loss}
\begin{figure}[htbp]
	\begin{center}
		\includegraphics[width=0.9\linewidth]{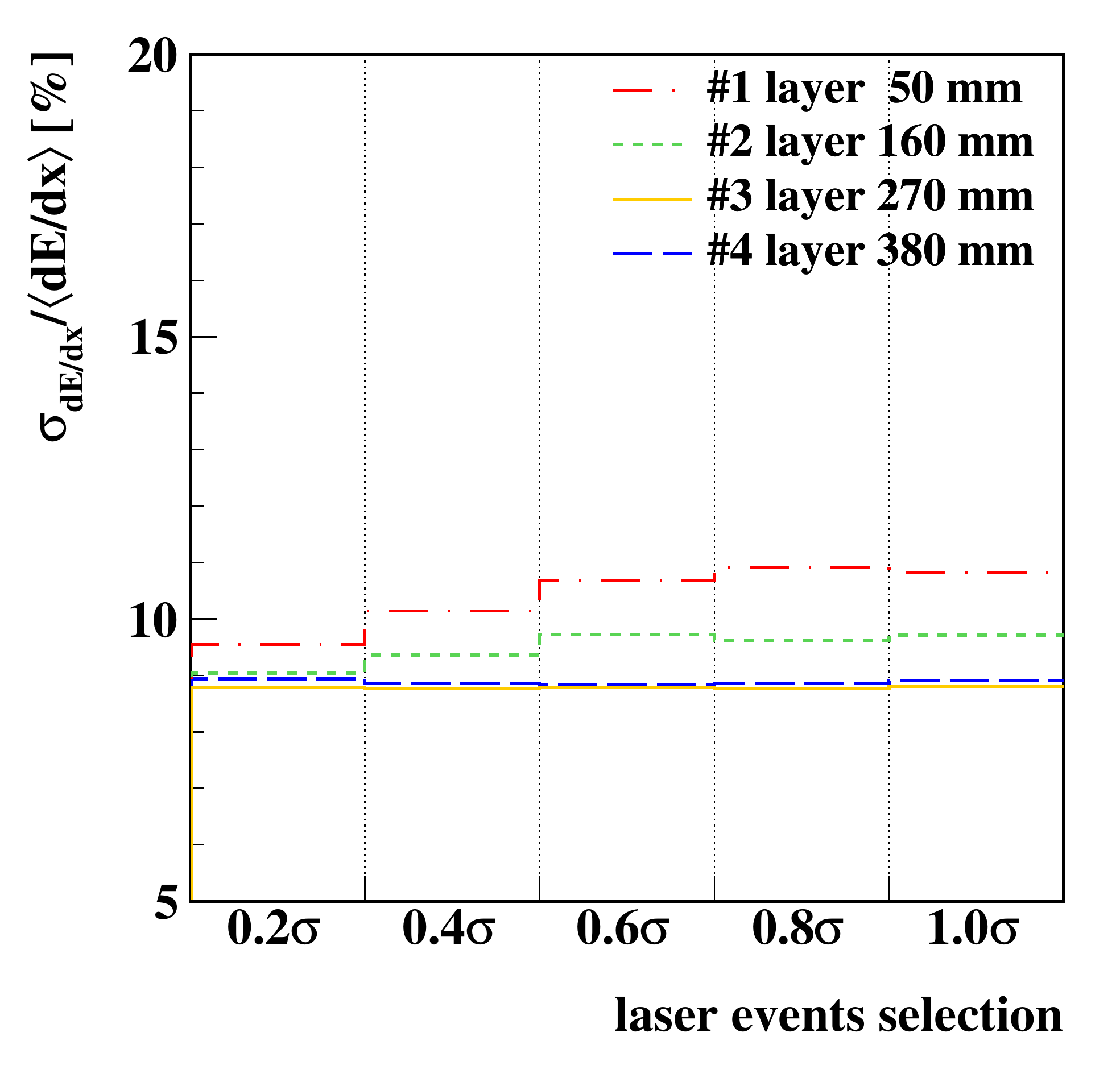}
		\caption{Comparison of dE/dx distribution estimated for different drift length $50\, mm$ (red line), $160\, mm$ (green line), $270\, mm$ (orange line) and $380\, mm$ (blue line).} 
		\label{Fig7}
	\end{center}
\end{figure}

In TPC detector, the energy loss is one of the means of particle identification, and for a particle with given momentum $p$ and mass $m$, its energy loss is a function about $\beta \gamma = p/(mc)$, where $\gamma$ is Lorentz factor. The TPC detector measures the energy loss $dE/dx$ by counting the amount of charge or number of electrons collected. There exists considerable fluctuations in energy loss, the mean value can be dominated by a few high energy deposits. To better estimate the mean value, the truncated mean method is utilized. Events with laser energies well above(below) the mean value are also excluded from the data analysis. A comparison of the method for different cut-off fractions for the laser energy is shown in Fig.~\ref{Fig7}. Different propagation paths in laser system lead to different energy loss. The measured ionization energy loss and its resolution will be slightly different. When calculating the transverse drift-diffusion, the laser energy of different layers is corrected to the same layer (the 4th layer) by multiplying the weighting factors.

\begin{figure}[htbp]
	\begin{center}
		\includegraphics[width=0.9\linewidth]{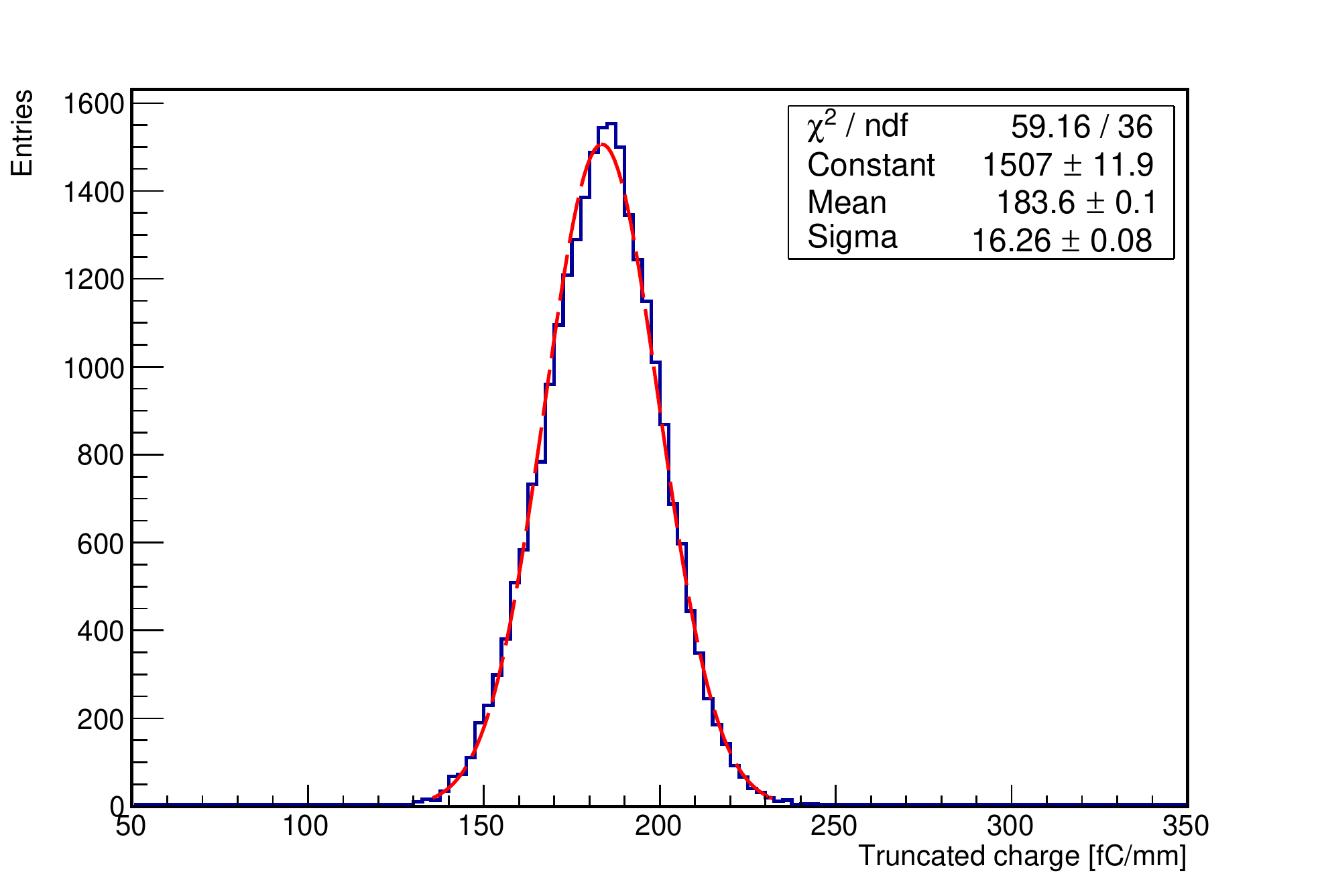}
		\caption{Distribution of the average $dE/dx$ of tracks. A Gaussian fit was applied in a $3\sigma$-range around the mean. The $\chi ^2/ndf = 59.16/36$ of the fit reflects the slight asymmetry visible in the distribution.} 
		\label{Fig8}
	\end{center}
\end{figure}

The distribution of the mean $dE/dx$ values, shown in Figure.~\ref{Fig8}, is almost Gaussian except for the slight asymmetry visible in the distribution. The $dE/dx$ resolution is determined to be $(8.9\pm 0.4)\, \%$ for events with 38 hits in a track.

\begin{figure}[htbp]
	\begin{center}
		\includegraphics[width=0.9\linewidth]{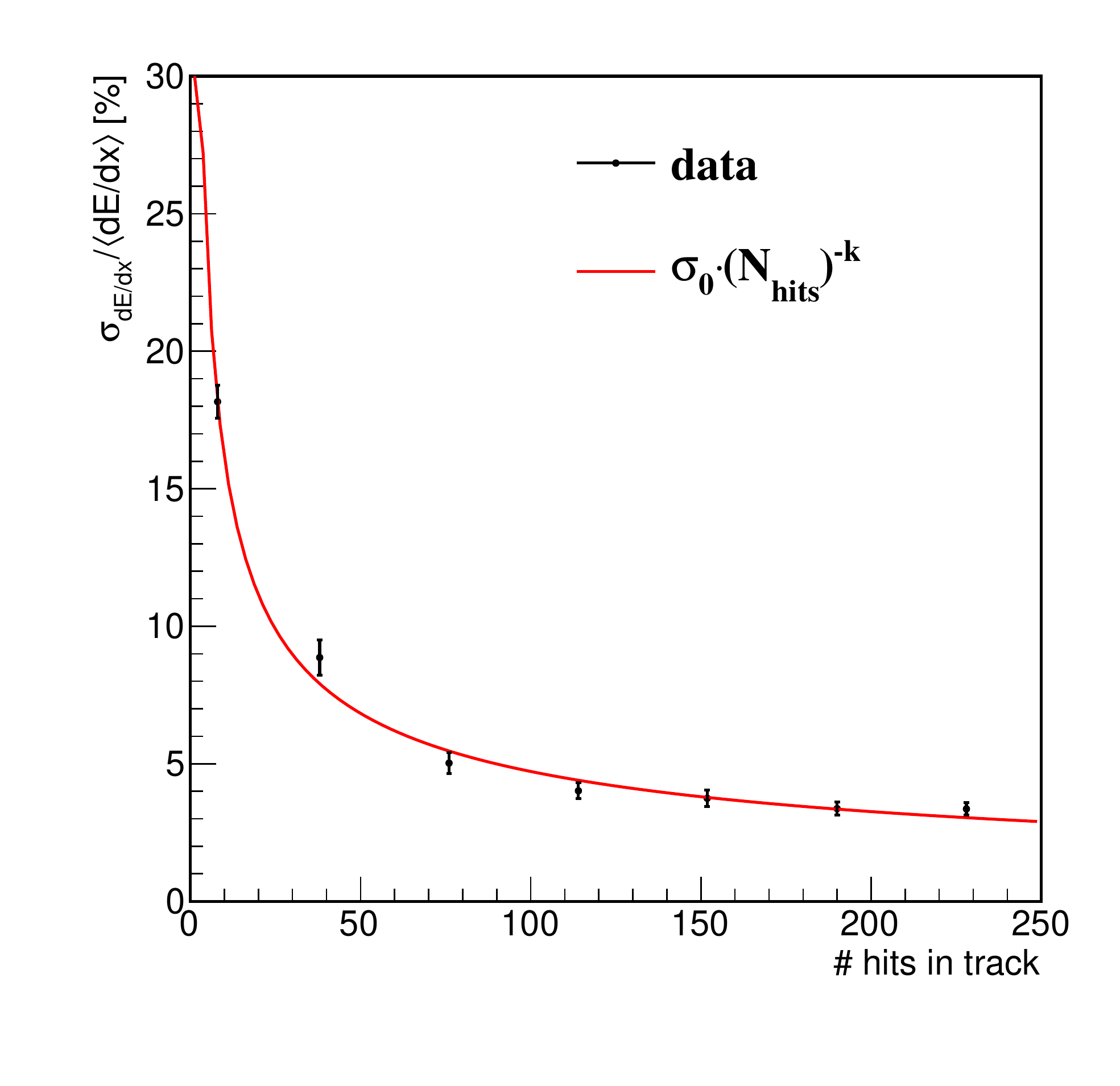}
		\caption{The dE/dx resolution versus the number of hits in a pseudo-track of various lengths. The fitted power law has $\chi ^2/ndf = 7.307/5$.} 
		\label{Fig9}
	\end{center}
\end{figure}

In order to estimate the performance of a full-scale 220-rings CEPC TPC, extrapolation of the prototype measurements is required. A power law\cite{einhaus2019studies} with an exponent $k\sim 0.5$ is expected, as given in Eq.~\ref{Eq3}.
\begin{equation}
	\sigma _{dE/dx}\propto {N_{hits}}^{-k},
\label{Eq3}
\end{equation}
where $N_{hits}$ is the number of hits in a track. Data of 46 columns of pad data is used for extrapolated energy loss performance study. Since the number of valid hits in the prototype is only 38, hits from multiple tracks in consecutive events are combined to form pseudo-tracks of arbitrary length, the result is shown in Fig.~\ref{Fig9}. The power-law fit returns an exponent $k = 0.53\pm 0.02$ and an intrinsic fluctuation for single hits $\sigma_{0} = 55.11\pm 3.36\, \%$. This leads to an estimate for the resolution at 220 hits of $\sigma_{dE/dx}=3.36\pm 0.26\,\%$. These values assume all hits are valid. Taking into account about $10\,\%$ invalid hits, result increases by about $0.6\,\%$.

\subsection{Drift velocity}
In TPC chamber, the gas property parameters are variable, partly from molecular dynamic, such as pressure and temperature changes, and partly from external factors, such as electric and magnetic field distortions. Changes in environmental factors can be corrected using long-time monitoring, and distortions in the electromagnetic field can be corrected using laser tracks in the good field region. 

\begin{figure}[htbp]
	\begin{center}
		\includegraphics[width=0.9\linewidth]{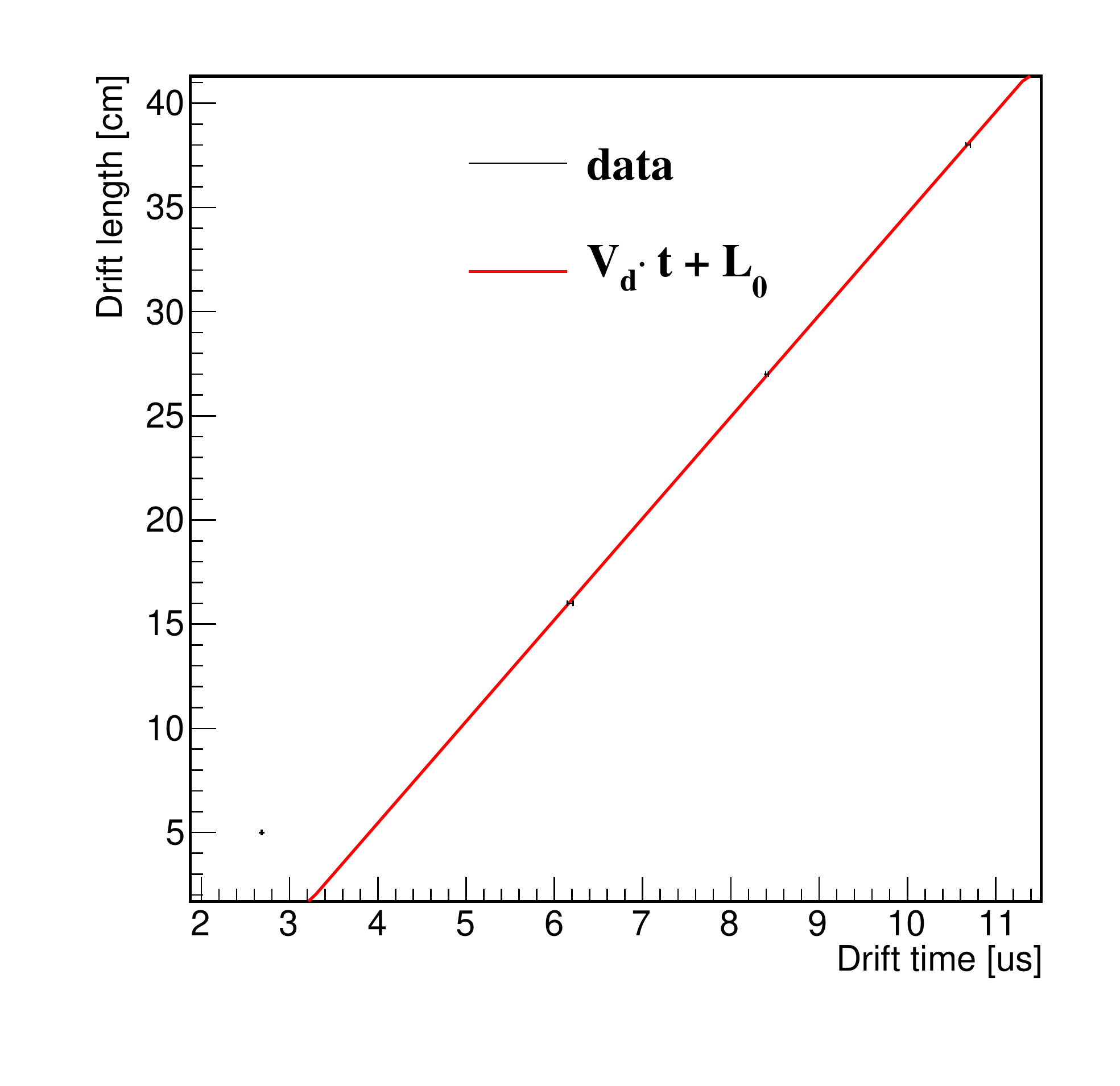}
		\caption{The drift time as a function of drift length. The fitted linear function has $\chi ^2/ndf = 0.9329/1$. The 1st data point is excluded in linear fitting.} 
		\label{Fig10}
	\end{center}
\end{figure}

For each column of pads perpendicular to the track direction, the method with caculating the charge center of gravity is used to reconstruct the drift time. It was found that the drift velocity of the first layer of laser closest to the upper surface of the first layer of GEM did not match the expected values compared to the drift velocity of the last three layers, as shown in Fig.~\ref{Fig10}. 

By utilizing the finite element analysis tool, the electric field depression phenomenon on the upper surface of the GEM was found, compared with the designed drift electric field of $180\,V/cm$. For the long drift, the measured drift velocity $(6.81\pm 0.15)\,cm/\mu s$ matches the expected value $7\,cm/\mu s$ from the simulation of Garfield++.

\section{Conclusion}
A system including a TPC prototype with 38 valid hits for each horizontal UV laser track, readout electronics with $200\times 200\,mm^2$ double GEM and the DAQ, was set up and tested with a laser system. The test results show that the hit resolution is limited essentially by effective number $N_{eff}$ and diffusion of electrons. The dE/dx resolution was measured to be $(8.9\pm0.4)\,\%$. Extrapolating this to the substantial track length, the resolution of the CEPC TPC (220 rings) is estimated at $(3.36\pm0.26)\,\%$, assuming a $100\%$ usable hits. This value has only a $0.6\%$ increment when considering a $10\%$ invalid hit rate. The calculation of velocity in the drift direction is affected by the distortion of the electric field, and the calculated value can be corrected to match the expected value by finite element analysis. 

If this prototype is used in the lepton collider, it can be a helpful device in the experiments. UV laser was the calibration device in ALICE TPC as an example project. Further studies on the high precision resolution requirements is needed in the future electron-positron colliders, especially CEPC project. This TPC prototype has been realized to confirm the function and set the narrow UV laser spot($0.5\,mm^2$) as the specific beam tracks. Indeed, our group has some plans to study more narrow laser spot, lower gain, and so on. 

\section*{Acknowledgement}
The author thanks for Prof. Yuanning Gao, Prof. Yulan Li and Dr. Yiming Cai for some details discussions. This study was supported by National Key Programme for $S\&T$ Research and Development (Grant NO.: 2016YFA0400400), the National Natural Science Foundation of China (Grant NO.: 11675197) and the National Natural Science Foundation of China (Grant NO.: 11775242)

\bibliography{mybibfile}

\end{document}